\pdfoutput=1
\documentclass[11pt]{article}
\usepackage{mathfin-paper}

\title{A Formally Verified Library of\\ Mathematical Finance in Lean~4}
\author{Raphael Coelho\thanks{Independent researcher. ORCID
  \href{https://orcid.org/0009-0001-6601-1023}{0009-0001-6601-1023}.
  Correspondence: \href{mailto:raphaelrrcoelho@gmail.com}{\texttt{raphaelrrcoelho@gmail.com}}.
  Artifact: \url{https://github.com/formal-applied-math/formal-mathfin}, release \texttt{v1.11.0}.}}
\date{November 2026}

\hypersetup{pdftitle={A Formally Verified Library of Mathematical Finance in Lean 4},
  pdfauthor={Raphael Coelho}}

\begin{document}
\maketitle

\begin{abstract}
\noindent
We describe a library of mathematical finance in the Lean~4 proof assistant, built on Mathlib and the
\texttt{BrownianMotion} package. Its benchmark records 804 results from the literature, each with a Lean
statement, a status, and a published account of what the formalization leaves out: 766 are proved in full,
18 restate an upstream theorem, and 20 are reduced cores that prove less than the result they are named
after. The library is about 158,000 lines of Lean, contains no \texttt{sorry}, and a build-time audit pins
every library constant that a benchmark proof cites to Lean's three standard axioms. The It\^o integral is
constructed rather than assumed. On it the library proves It\^o's formula for $C^2$ functions of It\^o
processes as an identity of processes, L\'evy's characterisation, martingale representation, Girsanov's theorem for bounded predictable drifts, which yields a Brownian motion for the filtration when the driver's paths are almost surely continuous, and unique
$L^2$ solutions of Lipschitz stochastic differential equations on every horizon. These results reach the
contracts a user prices. In a contract language that separates payoffs from models, the Black--Scholes call
is replicated by its delta hedge up to every date before maturity, and every fixed-dated contract on the
physical Black--Scholes asset has, under the Girsanov measure of a horizon containing its dates, its risk-neutral value. An audit of faithfulness makes the
logical status of every claim checkable, and its review has caught what mechanical gates cannot see,
including a convergence theorem whose hypothesis could never be satisfied. Nearly all of the mathematics is
classical, and what a formal base over it produces is certified unification of known results; two
contributed developments, by Robert Martin, address problems the library's own survey had listed as open.
The contribution is methodological and infrastructural.
\end{abstract}

\section{Introduction}

Interactive theorem provers have moved, over the last decade, from isolated landmark proofs to broad, reusable
libraries. Mathlib~\cite{Mathlib2020} now carries a substantial measure-theoretic probability development:
conditional expectation, martingales, the Radon--Nikodym theorem, Doob's theorems, Gaussian measures, and the
definition of pre-Brownian motion. Degenne and collaborators built Brownian motion itself on top of it,
in the \texttt{BrownianMotion} package~\cite{DegenneEtAl2025,BrownianMotionRepo}. Mathematical finance sits directly on this foundation: arbitrage-free pricing,
the Black--Scholes model~\cite{BlackScholes1973}, the fundamental theorem of asset
pricing~\cite{HarrisonKreps1979}, risk measurement. Little of it had been formalized.

What existed was narrow and deliberate. Echenim and Peltier formalized the binomial model and its completeness
in Isabelle/HOL~\cite{EchenimPeltier2017}, and with Guiol the Cox--Ross--Rubinstein pricing of European
derivatives~\cite{EchenimGuiolPeltier2020}. Pusceddu and Bartoletti formalized constant-product automated market
makers in Lean~4~\cite{PuscedduBartoletti2024}. Nagy gave a machine-verified derivation from It\^o's lemma to
Black--Scholes in Lean~4~\cite{Nagy2026}, a focused vertical that, on its author's account, leaves the continuous
It\^o integral structurally verified rather than constructed. More recently, Ushakov and Berdinsky priced European
and barrier options through a Henstock--Kurzweil path integral in Lean~\cite{UshakovBerdinsky2026}. Each of these
is self-contained, and none offers a way to state \emph{how faithful} a formalization is to the mathematics it
claims.

This paper describes a library that is broad, deep in its continuous-time foundations, and audited for
faithfulness. It covers thirteen areas, from the It\^o integral to quantitative risk management, and the state
reported here is release \texttt{v1.11.0} of November 2026. Its It\^o calculus and its fundamental theorems of asset
pricing are treated in detail in two companion papers~\cite{CoelhoIto,CoelhoFTAP}; this one describes the
library as a whole, the principles that keep it coherent, the path from its foundations to the contracts it
prices, and the audit that keeps it honest.

\paragraph{Changes from the third version}
The corpus has grown from 330 to 804 entries and the library from about 52,000 to about 158,000 lines. New since
the third version are, in continuous time, martingale representation and market completeness, It\^o's formula for
It\^o processes with continuous adapted coefficients as an identity of processes, the product rule, L\'evy's
characterisation, Girsanov's theorem for the filtration, stochastic differential equations on every horizon,
stochastic Fubini and the agreement of the integral with the \texttt{BrownianMotion} package's stochastic integral;
in discrete time, the reduction of the Dalang--Morton--Willinger theorem to its one-period conditional form and the
second fundamental theorem on finite spaces; and, for users, a contract layer on which the delta hedge and the change
of measure are theorems about contract values (\cref{sec:contracts}). Jump-diffusions, a quantitative-risk-management
benchmark, fixed income and credit have grown, and two contributed developments have been added
(\cref{sec:provenance}). Some results the library already contained, among them It\^o's formula under exponential
growth, are reported here for the first time. The audit now pins every cited constant rather than a curated list,
and the kernel replay, which could not run at the third version, completes. Several statements of the third version
are corrected here: the gloss of the reduced-core tier (\cref{sec:audit}), the list of instances of the Garman normal
form and the account of the survival identity (\cref{sec:architecture}), the status of the binomial limit, whose
stated hypothesis was unsatisfiable at the time (\cref{sec:vacuity}), and the description of the integral
$\int B\,\dd B$ (\cref{sec:depth}).

\section{Contributions}\label{sec:contributions}

\begin{enumerate}[label=(\arabic*)]
\item \emph{Breadth on one foundation} (\cref{sec:breadth}). To our knowledge the broadest formalization of
  mathematical finance in any proof assistant: thirteen areas and 804 catalogued results on one set of
  definitions, with cross-module reuse rather than per-instrument re-derivation.
\item \emph{Depth in continuous time} (\cref{sec:depth}). The It\^o integral is constructed as an isometry, and
  It\^o's formula, L\'evy's characterisation, martingale representation, Girsanov's theorem and stochastic
  differential equations are built on it; the risk-neutral measure is constructed by a change of measure rather than
  assumed.
\item \emph{Foundations that reach the contracts} (\cref{sec:contracts}). Payoffs are data and models are
  interchangeable, and the delta hedge, the change of measure and the multi-period fundamental theorem are stated
  as theorems about the value of a contract, not about a formula beside it.
\item \emph{A faithfulness audit} (\cref{sec:audit}), the most transferable contribution. Every result carries a
  tier that records how its Lean statement relates to the claim, and a published disclosure of what it leaves
  out. Build-enforced gates pin axioms, forbid unproved or definitional shortcuts in full entries, keep
  verification evidence fresh, and stop prose from naming an object a statement only asserts to exist.
  \Cref{sec:vacuity} reports what the gates missed and how it was found.
\item \emph{An honest map} (\cref{sec:evaluation,sec:limitations}). The reduced cores, the limitations of the
  headline results and of the apparatus itself, and the provenance of the code,
  including third-party ports and AI assistance (\cref{sec:provenance}).
\end{enumerate}

\section{Related work and positioning}\label{sec:related}

Formalized finance has appeared mostly at formal-methods venues. \Cref{tab:related} compares the nearest
precedents along the two axes the prior art leaves open. Nagy's derivation~\cite{Nagy2026} is the closest, and
the library builds on it directly: three modules (\file{Foundations/DiscreteIto.lean}, \file{FTAPTwoState.lean},
\file{ItoLemma.lean}) adapt its discrete It\^o, two-state fundamental theorem and structural-drift framework, with
attribution in their headers. Where Nagy machine-checks one It\^o-to-Black--Scholes path and stops at the
continuous It\^o integral, this library constructs that integral and sets the same derivation inside a
comprehensive, audited development. The work of Echenim, Guiol and Peltier~\cite{EchenimPeltier2017,EchenimGuiolPeltier2020}
is the established precedent for formalized pricing in discrete time. The library re-implements the
constant-product invariants of Pusceddu and Bartoletti~\cite{PuscedduBartoletti2024} for breadth, adapting their
mathematical content with attribution, and claims nothing new in automated market makers. Ushakov and
Berdinsky~\cite{UshakovBerdinsky2026} take a different analytic route to pricing, through the Henstock--Kurzweil path
integral rather than a stochastic integral.

\begin{table}[tbp]
\centering\small
\begin{tabularx}{\linewidth}{@{}>{\raggedright\arraybackslash}p{0.24\linewidth}>{\raggedright\arraybackslash}X>{\raggedright\arraybackslash}X>{\raggedright\arraybackslash}X@{}}
\toprule
 & this work & Nagy (2026) & Echenim et al.\ (2017--20) \\
\midrule
prover & Lean~4 & Lean~4 & Isabelle/HOL \\
scope & 13 areas, 804 results & one derivation (It\^o to Black--Scholes) & binomial and CRR models, completeness \\
continuous It\^o integral & constructed & structurally verified & not applicable (discrete time) \\
risk-neutral measure & derived by a change of measure & derived in a two-state market & discrete-time model \\
faithfulness audit & four tiers, per-entry disclosure, axiom gates & not part of the work & not part of the work \\
\bottomrule
\end{tabularx}
\caption{This work against the nearest formalized-finance precedents.}
\label{tab:related}
\end{table}

Mathlib and the \texttt{BrownianMotion} package are the upstream foundation. The library consumes Mathlib's measure
theory, conditional expectation, martingales and Gaussian machinery, and the package's construction of Brownian
motion, its simple processes and predictable $\sigma$-algebra, local martingales and axiomatic stochastic integral;
\cref{sec:evaluation} reports what has been contributed back. On the benchmark side, StochBench~\cite{StochBench2026}
collects 450 graduate stochastic-process problems in Lean and, lacking infrastructure such as the It\^o integral,
states 336 of them with the required properties as hypotheses. The library's benchmark is of a different kind: it
records results that are proved, and its reduced-core tier makes exactly that hypothesis-taking visible where it
remains.

\section{What the library proves}\label{sec:breadth}

\Cref{tab:areas,tab:areas2} list the thirteen areas with representative results. Each one-line statement describes the Lean
statement, not the textbook theorem it is named after; where they differ, the difference is given.

\begin{table}[tbp]
\centering\footnotesize
\renewcommand{\arraystretch}{1.1}
\begin{tabularx}{\linewidth}{@{}>{\raggedright\arraybackslash}p{0.15\linewidth}>{\raggedright\arraybackslash}X@{}}
\toprule
Area & Representative results \\
\midrule
Probability & Doob's $L^p$ maximal inequality for non-negative submartingales (\lean{maximal_ineq_Lp}); convergence of
  right-continuous $L^p$-bounded martingales, $p>1$, almost surely along the integers and in measure along the reals;
  the Ionescu-Tulcea path measure of a Markov chain on cylinders. \\
Stochastic calculus & The It\^o integral as an isometry (\lean{itoIntegralCLM}) and as a continuous local martingale; It\^o's
  formula without growth conditions, up to a continuous local martingale (\lean{ito_formula_unrestricted}), and for It\^o processes with
  continuous adapted coefficients as an identity of processes (\lean{ito_formula_adapted_unbounded_C2_process}); the product
  rule (\lean{ito_product_rule_process}); L\'evy's characterisation (\lean{isBrownianReal_of_localMartingale}); martingale
  representation; Girsanov's theorem for bounded predictable drifts (\lean{Btheta_isFilteredPreBrownian_predictable_of_bdd});
  Lipschitz SDEs on every horizon (\lean{picardMap_exists_unique_fixedPoint_any_horizon}); stochastic Fubini for any
  continuous linear integral (\lean{stochFubini_ofCLM}); the heat-equation link of Feynman--Kac~\cite{CoelhoIto}. \\
No-arbitrage & The fundamental theorem over several periods on a finite probability space of full support with one risky asset
  (\lean{ftap_discrete}) and, with full information at the horizon, the second theorem there (\lean{ftap_two}); in one period with finitely many assets, on a finite
  state space and on any probability space (\lean{ftap_one_period_vector})~\cite{CoelhoFTAP}; on any probability space,
  multi-period no arbitrage is equivalent to no arbitrage in each conditional one-period market
  (\lean{noArbitrage_iff_noArbitrageLocal}); an equivalent martingale measure excludes arbitrage by simple strategies in
  continuous time; change of num\'eraire; completeness of the Brownian market with a unique hedge
  (\lean{exists_replicating_strategy}). \\
Option pricing & The Black--Scholes call under a lognormal hypothesis that is itself derived from the physical model; its
  replication by the delta hedge (\lean{bs_delta_hedge_replication}); the Black--Scholes equation from Feynman--Kac
  (\lean{bsV_satisfies_bs_pde_via_feynmanKac}); Breeden--Litzenberger in second-order form for Black--Scholes and for
  jump-diffusions with a Gaussian part, and in first-order form for any law with a finite forward; jump-diffusions, assumed
  as a hypothesis structure: the discounted price is a martingale exactly at the compensated drift
  (\lean{JumpDiffusionProcess.martingale_iff}); with Gaussian log-jumps at that drift, Merton's 1976 series at every date
  before maturity (\lean{condExp_call_eq_mertonCallPrice}); Esscher pricing, and incompleteness at one date
  (\lean{exists_call_esscher_ne_merton}). \\
Exotic options & The $n$-date geometric Asian call in Black--Scholes form; Margrabe's formula as a Black--Scholes call on the
  price ratio when the ratio is lognormal; the chooser; quanto forwards; variance swaps, with realized variance converging to
  $\sigma^2T$ in $L^2$ for any drift; Stulz's rainbow and the Heynen--Kat barrier only at zero strike or zero correlation. \\
Lattice models & The Cox--Ross--Rubinstein call price converges to the Black--Scholes price
  (\lean{binomialPrice_call_tendsto_bs_closed}), with a lemma that its no-arbitrage hypothesis holds when
  $\abs r\sqrt T<\sigma$ (\lean{binomialNoArb_crr}); the tree's Snell envelope as a Mathlib supermartingale and American
  options as its value; Andr\'e's reflection principle as a bijection of lattice paths. \\
\bottomrule
\end{tabularx}
\caption{Representative results in probability, stochastic calculus and pricing. Every name is a Lean declaration in
the library.}
\label{tab:areas}
\end{table}

\begin{table}[!t]
\centering\footnotesize
\renewcommand{\arraystretch}{1.1}
\begin{tabularx}{\linewidth}{@{}>{\raggedright\arraybackslash}p{0.15\linewidth}>{\raggedright\arraybackslash}X@{}}
\toprule
Area & Representative results \\
\midrule
Fixed income, credit & The Vasicek bond price in affine form, with the integrated rate defined through its Wiener
  representation; the zero-coupon bond put for any integrated short rate in Wiener representation
  (\lean{valueDisc_bondPut_wiener}), with the Vasicek, Hull--White and piecewise-volatility GSR puts as instances; caps as
  sums of caplets; the forward
  measure as a change of num\'eraire; duration and convexity as price derivatives; hazard rates as logarithmic derivatives of
  survival; CDS protection as a contract paying at default; under independence, the first-to-default hazard as the sum of
  the single-name hazards; the KMV--Merton default probability through the distance to default. \\
Portfolio theory & CAPM from the equilibrium first-order condition; the Black--Litterman posterior mean as the unique solution of its
  precision-weighted normal equation, without the Bayesian derivation; Markowitz critical points
  from the Lagrangian; Kelly growth over $n$ periods; the Kelly portfolio as num\'eraire induces the risk-neutral measure in a
  two-outcome market. \\
Risk measures & The Artzner--Delbaen--Eber--Heath representation of coherent risk measures on a finite state space
  (\lean{coherentRisk_isLUB}); value-at-risk as the quantile of an arbitrary law; the Rockafellar--Uryasev formula, coherence
  and dual representation of expected shortfall for every integrable loss; elicitability of value-at-risk, on laws with a
  finite mean and a unique quantile, and its failure for expected shortfall. \\
Dependence, extremes & Sklar's theorem for continuous margins; Fr\'echet--Hoeffding bounds; max-stability of the GEV law;
  the generalized Pareto law in its domain of attraction; maxima of Poisson-many losses. Fisher--Tippett and
  Pickands--Balkema--de~Haan are not formalized. \\
Credit portfolios & The value-at-risk of the one-factor Gaussian default fraction converges to the asymptotic
  single-risk-factor quantile, the worst-case default rate at the core of the Basel IRB formula
  (\lean{tendsto_valueAtRisk_oneFactorLossFraction}). \\
Microstructure & A positive Glosten--Milgrom spread from trader-mix primitives; for one asset, the quadratic ansatz of
  Bergault--Evangelista--Gu\'eant--Vieira solves the approximate (linear--quadratic) Hamilton--Jacobi equation whenever its
  coefficients solve the associated ordinary differential equations, and for several assets the matrix Riccati equation of
  the leading coefficient is verified; the true value function is out of scope. \\
Other & A contract language that separates payoffs from pricing models, after Bilokon~\cite{Bilokon2026}
  (\cref{sec:contracts}); survival models, annuity-certain closed forms and compound Poisson losses; constant-product AMMs;
  strict convexity of the American put's exercise boundary and a polynomial-sign characterisation of Durrleman's condition
  for raw SVI, both by Robert Martin, the latter extended by the library to butterfly-free smiles (\cref{sec:provenance}). \\
\bottomrule
\end{tabularx}
\caption{Representative results in the applied areas. Every name is a Lean declaration in the library.}
\label{tab:areas2}
\end{table}

No individual theorem in the table is new; the mathematics is classical throughout, with the exception of the two
contributed developments discussed in \cref{sec:provenance}. The contribution is the development as a whole: one
foundation under all of it, depth where earlier efforts assume, and an audited statement of what each result proves.

\section{Depth: the continuous-time spine}\label{sec:depth}

The library's depth comes from carrying stochastic integration through to a genuine It\^o integral, the foundation
earlier formalizations assume or sketch, and then consuming it. The companion paper~\cite{CoelhoIto} develops this
in full; here is the spine.

\paragraph{The integral}
The It\^o integral is a linear isometry from the square-integrable predictable integrands on $(0,\infty)\times\Omega$
into $L^2(\mu)$ (\lean{itoIntegralL2}), the unique continuous extension of the elementary integral on simple adapted
processes, obtained by a $\pi$--$\lambda$ density argument and Mathlib's \lean{LinearMap.extendOfNorm}. On a horizon $T$,
$\ito_T$ is the same integral applied to integrands extended by zero beyond $T$ (\lean{itoIntegralCLM}), so there is one
construction, not one per horizon. As a process it is
the conditional projection of its terminal value, hence a continuous $L^2$ martingale, and it has a continuous
modification that is a local martingale on the half-line. For a continuous adapted integrand with no bound the integral
is built by gluing clipped integrals, and has a version that is a continuous local martingale. Its range together with
the constants is all of $L^2(\F^B_T)$, which is martingale representation, and it satisfies the \texttt{BrownianMotion}
package's axiomatic characterisation of a stochastic integral. One worked example is the identity behind
$\int_0^TB\,\dd B$: an integrand $g_B$, constructed as the limit of truncated step processes of $B$, has
$\ito_T(g_B)=\frac12(B_T^2-B_0^2-T)$ (\lean{itoIntegralCLM_brownian}). That statement does not itself name $g_B$ as $B$; the
named form is an instance of It\^o's formula under exponential growth, which does name its integrand, though that
instance is not stated separately~\cite{CoelhoIto}.

\paragraph{It\^o's formula}
In $L^2$, for $f(t,x)$ whose partial derivatives up to $f_{tt}$, $f_{tx}$ and $f_{xxx}$ are bounded or grow at most like
$e^{\lambda\abs x}$, $f(T,B_T)-f(0,B_0)=\ito_T(f_x(\cdot,B))+\int_0^T(f_t+\frac12f_{xx})(s,B_s)\,\dd s$ with the integrand
named. For every $C^2$ function, with no growth condition, the compensated process agrees, at each time almost surely, with a continuous local martingale on the null-augmented filtration, obtained by localization at exit times (\lean{ito_formula_unrestricted}). For an It\^o process
$X=x_0+\int b\,\dd s+\int\sigma\,\dd B$ whose coefficients are adapted, with every path continuous and no bound, and for
$f\in C^2$, almost surely and for every $t\le T$ at once,
$f(X_t)=f(x_0)+\int_0^t(f'(X_s)b_s+\frac12f''(X_s)\sigma_s^2)\,\dd s+\int_0^tf'(X_s)\sigma_s\,\dd B_s$
(\lean{ito_formula_adapted_unbounded_C2_process}), and for time-dependent $f(t,x)$ the compensated process is a continuous
local martingale. Applied to $(X+Y)^2$, $X^2$ and $Y^2$ for two It\^o processes driven by the same Brownian motion, this
gives the product rule $\dd(XY)=X\,\dd Y+Y\,\dd X+\sigma\tau\,\dd t$ as an identity of processes
(\lean{ito_product_rule_process}). The general two-dimensional formula remains a reduced core.

\paragraph{L\'evy's characterisation}
An adapted process with continuous paths, started at zero, such that $M$ and $M^2-t$ are local martingales is a Brownian
motion (\lean{isBrownianReal_of_localMartingale}) whose increments are independent of the past
(\lean{isFilteredPreBrownian_of_localMartingale}). The bracket enters as the
hypothesis on $M^2-t$; no quadratic-variation process is constructed.

\paragraph{The risk-neutral measure, derived}
In most treatments, and in earlier formalized pricing, an equivalent martingale measure is posited and prices are its
expectations. Here it is constructed. A static Girsanov change of measure by an Esscher density turns the physical
Gaussian driver into one under which the terminal asset has exactly the lognormal law the Black--Scholes layer takes as
its hypothesis, so that hypothesis is a theorem (\lean{BSCallHyp.exists_of_physical}) and the call formula holds under a
measure constructed from the physical Gaussian driver alone (\lean{bs_call_formula_of_physical}). The dynamic Girsanov
theorem is proved along a ladder of drifts (constant, simple adapted, continuous adapted, bounded predictable). At the top
of the ladder, under $\frac{\dd Q}{\dd\mu}=\exp(-\ito_T(\theta)-\frac12\int_0^T\theta^2\,\dd s)$, the process
$B+\int\theta\,\dd s$ is a pre-Brownian motion for the natural filtration of $B$ on the whole half-line: it starts at zero,
and each increment is $N(0,t-s)$ and independent of $\F^B_s$ (\lean{Btheta_isFilteredPreBrownian_predictable_of_bdd}).
When the paths of $B$ are almost surely continuous, so are those of $B+\int\theta\,\dd s$ under $Q$, which is then a
Brownian motion (\lean{Btheta_isBrownianReal_predictable_of_bdd}). The third version proved only the finite-dimensional
laws. The discounted geometric Brownian motion is an It\^o integral with its integrand named
(\lean{discountedGBM_eq_itoIntegral}), so the pricing layer meets the continuous world through the integral itself.

\paragraph{Completeness and its limits}
For a Brownian motion with every path continuous, every square-integrable $\F^B_T$-measurable claim is
$\E[H]+\int_0^T\varphi\,\dd B$ for a unique hedge (\lean{exists_replicating_strategy}); integrating against the price $S=S_0+\int\sigma\,\dd B$ instead, with $\sigma\ne0$
almost everywhere, the hedge is a holding in the price. A probability measure with square-integrable density under
which $S$ is a martingale agrees with $\mu$ on $\F^B_T$: the direction of the second fundamental theorem from
completeness to uniqueness. The converse, the general Girsanov theorem under Novikov's condition, and the identification
of the hedge (Clark--Ocone) are open.

\paragraph{Equations and jumps}
For Lipschitz coefficients the Picard map of $\dd X=b(X)\,\dd t+\sigma(X)\,\dd B$ has exactly one fixed point in the
predictable $L^2$ space on every horizon (\lean{picardMap_exists_unique_fixedPoint_any_horizon}): some iterate is a
contraction, with factorial decay, so no smallness condition on $T$ is needed. On short horizons, where
$TL_b+\sqrt TL_\sigma<1$, the fixed point is realized as a sample-path process (\lean{sde_pathwise_decomposition}).
Stochastic Fubini is stated once, for any continuous linear map out of $L^2$, with the Wiener and It\^o integrals as
instances (\lean{stochFubini_ofCLM}). Jump-diffusions enter through a hypothesis structure (\lean{JumpDiffusionProcess}): an
adapted process with independent increments whose laws are those of drift plus Gaussian plus compound Poisson. On it the
discounted price is a martingale exactly at the compensated drift, with Gaussian log-jumps at that drift the conditional call price at every date before maturity is Merton's
series~\cite{Merton1976}, and, for a non-degenerate jump law with exponential moments, the Esscher measure and the
compensated Merton law price some call differently at one date. Such a process is
constructed only without jumps, as Brownian motion with drift.

\paragraph{From lattice to continuum}
The Cox--Ross--Rubinstein call price converges to the Black--Scholes price, by characteristic functions and L\'evy's
continuity theorem, with put--call parity removing the uniform-integrability obstruction. The limit is proved for call
prices and for the law of the terminal log-return; Donsker's invariance principle is not formalized.

\section{From foundations to contracts}\label{sec:contracts}

A practitioner prices contracts, not theorems about Brownian motion. The library's contract layer
(\file{Contracts/}) makes payoffs data: a \lean{Contract} is built from observations of named underlyings at fixed or
random dates, by combinators, and \lean{Contract.value} prices it as the expectation, under a chosen measure, of its
discounted pathwise present value against a model that turns each observation into a random variable. The design follows
Bilokon's \emph{The Contract Is Not the Model}~\cite{Bilokon2026}; no code is taken from it. One contract term is priced in
the Black--Scholes model, the binomial tree, Gaussian short-rate models with a stochastic discount factor, and a discrete
market on any probability space, and the term is the same in each. For the European call, put, digital and capped call
under single-asset Black--Scholes the values reduce to the closed forms the pricing layer proves, and the call's value in the $n$-step tree converges to its value under the lognormal hypothesis \lean{BSCallHyp} when
$\abs r\sqrt T<\sigma$ (\lean{tendsto_binomial_value_europeanCall}).

The foundations of \cref{sec:depth} reach this layer through theorems about contract values.
\begin{itemize}
\item \emph{The delta hedge.} Let $S_t=S_0e^{(r-\sigma^2/2)t+\sigma B_t}$ be the risk-neutral Black--Scholes path, $C(t,S)$
  the call price and $T'<T$. Almost surely, for every $t\le T'$ at once,
  $e^{-rt}C(t,S_t)=V_0+\int_0^t\Phi(d_1)\,\sigma e^{-rs}S_s\,\dd B_s$, where $V_0$ is the value of the call contract
  (\lean{europeanCall_deltaHedge}). Undiscounted, the same identity is the self-financing equation
  $V_t=V_0+\int_0^t\Delta_s\,\dd S_s+\int_0^tr\psi_s\,\dd s$ with bank position $\psi=V-\Delta S$
  (\lean{europeanCall_selfFinancing}), and along every path continuous at $T$ the discounted call price tends to the
  discounted payoff of the contract as $t\uparrow T$ (\lean{tendsto_hedge_europeanCall_payoff}). The path solves its
  stochastic differential equation, $S_t=S_0+\int_0^trS_s\,\dd s+\int_0^t\sigma S_s\,\dd B_s$ (\lean{pathAssets_solves_sde}).
\item \emph{Physical value is risk-neutral value.} For $\sigma\ne0$ and every fixed-dated contract observed in $[0,H]$, its
  value on the physical asset $S_0e^{(m-\sigma^2/2)u+\sigma W_u}$ under the Girsanov measure of $\theta=(m-r)/\sigma$ on $[0,H]$ equals its
  value on the risk-neutral asset driven by $W+\theta u$ (\lean{value_physical_eq_riskNeutral}). The physical drift enters
  only through the measure. The call, the chooser and the geometric Asian call are instances.
\item \emph{Hedging in the tree.} In the complete binomial tree, a European contract paying a function of the spot at a
  grid date is replicated by a bounded predictable strategy whose initial capital is the contract's value
  (\lean{exists_hedge_eq_value}). The strategy is shown to exist; no delta formula is given.
\item \emph{No arbitrage on any probability space.} For a one-asset discrete market, no arbitrage yields one equivalent
  martingale measure under which every forward and every call--put pair priced as contracts satisfy parity, assuming the
  conditional one-period fundamental theorem (\lean{exists_isEMM_parity}). The reduction behind it is proved without
  assumptions: multi-period no arbitrage is equivalent to no arbitrage in each conditional one-period market
  (\lean{noArbitrage_iff_noArbitrageLocal}), and bounded positive one-period densities paste into an equivalent martingale
  measure (\lean{isEMM_pastedMeasure}). The remaining step, which produces those densities by measurable selection, is
  carried as the hypothesis \lean{ConditionalOnePeriodFTAP}, and the entries that use it are reduced cores.
\item \emph{Fixed income.} Bond puts, caplets and coupon bonds are contract values under a stochastic discount factor in
  the Vasicek, Hull--White and piecewise-volatility GSR models. In an HJM model, fixed cash flows are worth $\sum_iP(0,t_i)c_i$
  given a deterministic initial curve and the hypothesis that the discount factor prices each bond
  (\lean{valueDisc_fixedFlows_hjm}); the drift condition that would deliver that hypothesis is not proved, and the entry is
  a reduced core.
\end{itemize}
Not every foundation reaches the contracts yet. The product rule, stochastic Fubini and the fixed point on every horizon
are proved, but no contract theorem uses them: the self-financing equation is derived from the delta-hedge identity, not
from the product rule, and the HJM bond theorem does not use Fubini.

\section{Architecture: principles and bridges}\label{sec:architecture}

The library stays coherent by deriving results from a few reusable principles rather than re-proving each instrument.
Four pillars carry the mathematics, and a small set of bridges connects them.

\paragraph{I. No-arbitrage as convex duality}
The finite fundamental theorem is a separating hyperplane between the attainable gains and the simplex; the
Artzner--Delbaen--Eber--Heath representation of coherent risk measures~\cite{ADEH1999} separates a rejected position from
the acceptance cone. The two separations are sibling lemmas in one module, proved from the same two helper
lemmas~\cite{CoelhoFTAP}. One equivalent-martingale-measure predicate serves every time index, discrete and continuous,
and one replication argument shows, in the finite market and in the Brownian market alike, that completeness makes the
pricing measure unique.

\paragraph{II. Stochastic calculus}
Continuous-time models are built on Brownian motion, and the It\^o integral and formula of \cref{sec:depth} are the
calculus for their functionals; pricing hypotheses stated on marginal laws are tied to Brownian motion by the bridges
below.

\paragraph{III. The probabilistic--analytic duality}
A price is both a risk-neutral expectation and the solution of a partial differential equation, and Feynman--Kac is the
bridge. The Black--Scholes equation is derived by writing the price as a heat-kernel integral and using the heat equation
for the kernel, without differentiating the closed form.

\paragraph{IV. Exponential families and intensities}
Closed forms come from Gaussian, lognormal and exponential structure. One Esscher module serves three users: its tilt,
Mathlib's \lean{Measure.tilted} at a linear exponent, gives the static Girsanov change of measure and Esscher pricing of
jump-diffusions, and its moment-generating-function uniqueness lemma identifies Brownian increments under a new measure.
And one exponential of an integrated intensity, \lean{survivalFromIntensity}, is the shared definition of the
intensity-based actuarial and credit survival functions.

\paragraph{The Garman normal form}
Many Black--Scholes-family prices have the form $A\,\Phi(d_1)-K\cdot\mathrm{DF}\cdot\Phi(d_2)$. The library proves this
equality for the standard call, Black-76 futures options, payer swaptions, the call with a continuous dividend yield, and
Margrabe's exchange option~\cite{Margrabe1978}. The third version listed the put,
digital and power options among the proved instances; they are not. Garman--Kohlhagen is an instance only through the
dividend-yield case, without a theorem of its own.

\paragraph{Grounding in Brownian motion}
Pricing hypotheses are stated marginally (\lean{BSCallHyp} asks only that the terminal asset be lognormal under the pricing
measure), and bridge modules show that they follow from a pre-Brownian motion, through which the flagship prices connect
to the continuous-time object.

\paragraph{Certified unification}
What a formal base over classical mathematics produces is not new theory but machine-checked identities between objects
developed for different purposes. For a Gaussian vector of losses, over portfolio weights with a common expected loss, minimising value-at-risk or
expected shortfall at a level above one half is equivalent to minimising variance
(\lean{isMinOn_valueAtRisk_iff_isMinOn_portfolioVarN}, \lean{isMinOn_expectedShortfall_iff_isMinOn_portfolioVarN}). The
$N$-asset portfolio variance with uncorrelated assets of common variance is that variance times the Herfindahl concentration
index (\lean{portfolioVarN_diag_eq_herfindahl}), and with equal correlation it is an explicit mixture of the index and the
squared total weight. Mortality and reduced-form credit share one intensity-based survival function. The third version presented this last identity
as one between independently developed definitions; in fact both are defined through a shared
\lean{survivalFromIntensity}, so it holds by definition, and it is the shared definition, not the identity, that records the
unification.

\paragraph{One graph}
Coherence is also enforced. Foundations import only foundations, and joins live in bridge modules; every module is
imported by another or cited by a benchmark entry, so nothing sits disconnected. A declaration index, built from the Lean
environment on every push, records each declaration's statement, and a ratchet fails the build if the number of duplicate
statements or of private lemmas whose statements mention no library constant rises; both counts are zero. Before the
release described here an audit of the library against these rules removed about sixty parallel developments, among them
five definitions of the Riemann--It\^o sum and two forward measures, each folded into the library's own framework without
leaving an alias.

\section{The faithfulness audit}\label{sec:audit}

Formalization papers often leave implicit the gap between ``we proved $X$'' and ``we proved a weaker $X$ under added
hypotheses''. The library makes the gap explicit and partly machine-checkable.

\subsection{Tiers and disclosures}
Every benchmark entry declares one of four statuses. \status{full}: the Lean statement is the mathematical claim, derived
from foundational primitives. \status{library\_wrapper}: a thin restatement of an upstream Mathlib or \texttt{BrownianMotion}
theorem. \status{reduced\_core}: an honest but narrower statement. Of the twenty current cases, eleven are structure
specifications, in which the textbook conclusion is a field of a structure and the theorem reads it off by projection, so
nothing is derived; three prove the theorem from a named hypothesis that is the open step (the conditional one-period
fundamental theorem, twice, and the HJM bond-pricing hypothesis); four concern transcribed closed forms of exotic options,
proved in a special case or through their in--out parities; and two prove a special case. \status{placeholder}: none
remain. Delivery claims count only \status{full} and \status{library\_wrapper}. (The third version glossed the reduced
core as ``holds under an added hypothesis or with an axiomatized sub-step'', which understated how little the structure
specifications prove.) Each entry's \texttt{description} states what the entry proves, not the textbook theorem it is named
after, and its \texttt{formalization\_scope} gives the long-form disclosure of what it leaves out. Both are published,
together with the Lean statement, in a public dataset~\cite{HFDataset}.

\subsection{Machine-checked gates}
The tiers are metadata, checked by test gates that continuous integration runs before the Lean build, and the Lean build
itself checks axioms.
\begin{itemize}
\item \emph{Axiom audits.} A curated file pins the \texttt{\#print axioms} output of 411 headline constants under
  \texttt{\#guard\_msgs}. A generated file pins 958: every library constant cited in proof position by any benchmark
  snippet (929), resolved by declaration through \texttt{open}s, dot notation and foreign namespaces, and the 29 upstream
  constants that wrapper entries re-export. Together they pin 1{,}111 distinct constants. The upstream pins matter because
  \texttt{BrownianMotion} contains \texttt{sorry}s at the pin; none lies beneath a pinned constant. Any axiom beyond
  \texttt{propext}, \texttt{Classical.choice} and \texttt{Quot.sound}, any \texttt{sorryAx}, or a renamed constant fails the
  build.
\item \emph{Text gates.} After comment stripping, library sources contain no \texttt{sorry}, \texttt{admit},
  \texttt{native\_decide}, \texttt{polyrith}, search-tactic suggestions or premise-selection calls.
\item \emph{Shape gates.} A full entry may not bundle an assumed \texttt{Prop}-valued structure (the reduced-core pattern),
  nor rest on a definitional \texttt{rfl} proof; every reduced-core description must say what is not derived; and an
  existential statement whose witness is never pinned down may not be described by prose that names an integral of it.
\item \emph{Evidence freshness.} A verification ledger records, for every entry, a hash of its snippet, the transitive library
  modules it imports and the toolchain pins under which it last verified; an entry whose inputs change becomes stale and must
  re-verify.
\end{itemize}

\subsection{What the gates cannot see}\label{sec:vacuity}
Two failure modes pass every mechanical gate. The first is vacuity. The Cox--Ross--Rubinstein convergence theorems
that the third version of this paper reported as checked end to end carried the hypothesis that the $n$-step tree is
arbitrage-free for every $n$,
which fails at $n=0$, where the step size is $T/0=0$; so the theorems were true of no market and said nothing. No axiom
audit, kernel replay or \texttt{sorry} gate can detect this. It was found by reading the statement, not by any gate,
and fixed by quantifying over $n\ge1$ and proving that the hypothesis then holds whenever
$\abs r\sqrt T<\sigma$ (\lean{binomialNoArb_crr}). The external check of \cref{sec:external} likewise pairs its
theorem with a satisfiability witness.

The second is prose that outruns its statement: the build is green precisely because the statement is weaker than the
description. The library has repeatedly found such drift in its own text, including the Girsanov results, once described as
producing a Brownian motion when they fixed only finite-dimensional laws, and the integral $\int B\,\dd B$ of
\cref{sec:depth}. Only its mechanical slice is gated. The rest is a review discipline: every change to proof content
closes with a multi-agent review against eight lenses, whose standing first pass reads every touched description, docstring
and paragraph against its statement, and whose cadence (one review per twelve new entries at most) is itself test-enforced.
This paper was revised with the same first pass; the corrections listed in the introduction are what it found.

\subsection{External checks}\label{sec:external}
For the coherent-risk representation theorem the repository carries a Mathlib-only restatement, with a witness that coherent
risk measures exist, and a proof of both from the library. Lean's comparator~\cite{Comparator} checks, on every change to
them and monthly, that the proved statements are exactly the restated ones, that only the three standard axioms are used, and,
in its paranoid mode, that every external kernel bundled with Lean accepts the proof, not only Lean's own. A kernel replay of
every content module through \texttt{leanchecker}, below the elaborator, runs monthly in six parallel shards. It could not
finish at the third version, when every module imported all of Mathlib; since the imports were narrowed it completes.

\section{Evaluation}\label{sec:evaluation}

\begin{table}[tbp]
\centering\small
\begin{tabular}{@{}lr@{}}
\toprule
Benchmark entries & 804 \\
\quad \status{full} & 766 \\
\quad \status{library\_wrapper} & 18 \\
\quad \status{reduced\_core} & 20 \\
\quad \status{placeholder} & 0 \\
Delivery-ready (\status{full} $+$ \status{library\_wrapper}) & 784 \\
\midrule
Lean files / lines under \file{MathFin/} (excluding audit and blueprint files) & 722 / 158{,}240 \\
\quad of which contributed ports (\cref{sec:provenance}) & 213 / 27{,}518 \\
Theorems in the declaration index (public / private) & 5{,}468 / 325 \\
Constants pinned by the axiom audits (curated / generated / distinct) & 411 / 958 / 1{,}111 \\
Entries with a fresh verification-ledger row & 804 \\
\bottomrule
\end{tabular}
\caption{The library at release \texttt{v1.11.0}.}
\label{tab:numbers}
\end{table}

\Cref{tab:numbers} gives the numbers, from the repository's coverage report and source tree. The structure-specification
cores sit at the genuinely hard frontier of continuous-time and discrete-process theory: the reflection principle, nowhere
differentiability and the law of the iterated logarithm for Brownian motion; Novikov's condition, the general Girsanov
theorem and the two-dimensional It\^o formula; and five theorems on Markov chains (the strong Markov property, recurrence
criteria, uniqueness of the stationary distribution, the ergodic theorem and convergence to equilibrium). L\'evy's
characterisation, a reduced core at the third version, is now proved. The other reduced cores are the three conditional
results of \cref{sec:contracts}, four exotic closed forms, the exponential law of the first interarrival time of a Poisson
process, and the decay of the Ornstein--Uhlenbeck Hermite modes up to degree two. The entries in mathematical finance, the
largest domain, are 657 full of 666. The library is most faithful where the mathematics is classical and elementary, and
its partial cores sit where the infrastructure is missing.

\paragraph{Reproducibility}
The toolchain is pinned: Lean \texttt{v4.35.0}~\cite{deMouraUllrich2021}, Mathlib at \texttt{e91b128} and
\texttt{BrownianMotion} at \texttt{497e5e2}, in \file{lean-toolchain} and \file{lake-manifest.json}. A plain \texttt{lake build}
from the repository root checks every proof; release \texttt{v1.11.0} is archived on Zenodo~\cite{Zenodo}, and a pinned container
image (\texttt{ghcr.io/formal-applied-math/mathfin-verify}) reproduces the build.

\section{Provenance}\label{sec:provenance}

A library that audits its claims should also account for its code. The repository records provenance file by file and in a
machine-readable self-report.

\paragraph{Contributed developments}
About a sixth of the source lines are ports of two developments by Robert Martin, each kept under its original authorship. The first
proves, in the constant-parameter Black--Scholes model with $r>0$ and dividend yield $0\le q\le r$, that the American put's
early-exercise boundary $B(\tau)$ is strictly convex in the time to expiry and that $\log(B(\tau)/K)$ is
convex~\cite{MartinAmericanPut}. The value is the supremum over all stopping times of the completed Brownian filtration, and the
headline theorems assume no classical solution of the free-boundary problem; a further, conditional part of the port, stated over
an assumed classical solution that nothing shows inhabited, is not claimed. The second gives a fixed Boolean combination of signs
of polynomials in the five raw-SVI parameters that holds exactly when the parameters are admissible and Durrleman's condition
holds, that is, when the call priced through the SVI smile is convex in the strike~\cite{MartinSVI,GatheralJacquier2014}. The
library adds the large-strike tail condition: with one further sign the formula holds exactly when that call is a normalised call
price function in Roper's sense, which is freedom from butterfly arbitrage~\cite{Roper2010}; recovering the risk-neutral law from
such a function is not formalized.
The repository's survey of open problems had listed both questions as open, the first for $0<q<r$ and the second as an explicit
semialgebraic description of the SVI domain, and records both as resolved by these developments. The credit is Robert Martin's;
the first was, on its author's report, substantially AI-assisted; neither has been refereed; and their status is that of
machine-checked, axiom-pinned results.

\paragraph{Other sources}
Nagy's derivation~\cite{Nagy2026} and the automated-market-maker formalization of Pusceddu and Bartoletti~\cite{PuscedduBartoletti2024}
are adapted in four modules with attribution headers. The contract layer takes design ideas, but no code, from
Bilokon~\cite{Bilokon2026}. Yosuke Ito's Isabelle entry on actuarial mathematics was consulted, with permission, for the survival
layer. Yang Ye contributed results across fixed income, credit, risk and exotic options, among them CVA, the
large-homogeneous-portfolio tranche loss, inflation swaps, the bivariate normal distribution with Stulz's and Heynen--Kat's
two-asset options, SA-CCR and decentralised-finance mechanisms. Alfredo Garcia contributed the Glosten--Milgrom model, the drifted
price path and the closed-form CIR loadings, Aleksey Salkutsan downside performance metrics and forward-rate agreements, and Mert
Unsal the Black--Scholes charm. Each module's header names its authors. Much of the benchmark follows Saporito's lecture notes on
stochastic processes~\cite{Saporito}, the corpus spine.

\paragraph{AI assistance}
The library was written with AI assistance, and the repository says so. The author used an AI coding assistant (Claude Code) as an
authoring assistant throughout. Four benchmark entries came from an automated pipeline in which language models drafted and proved the
statements; each was reviewed and refined by hand, and the pipeline is treated as a scout, not an author. The gates of
\cref{sec:audit} apply to all code unchanged, whoever or whatever wrote it, and every proof is checked by the Lean kernel when the
library builds.
This manuscript was revised with the same assistant, and its statements were checked against the Lean sources of release
\texttt{v1.11.0}.

\section{Limitations}\label{sec:limitations}

The limitations of the headline results are those their statements carry.
\begin{itemize}
\item Girsanov's theorem is proved for bounded predictable drifts; Novikov's condition is not.
\item In continuous time only the direction of the second fundamental theorem from completeness to uniqueness is proved,
  and martingale representation proves that a hedge exists and is unique without identifying it.
\item It\^o's formula for an It\^o process needs every path of the driver and of the coefficients to be continuous. The
  general two-dimensional formula is a reduced core; the product rule is proved.
\item The solution of a stochastic differential equation on a long horizon is an $L^2$ class, with no continuous version or
  pathwise uniqueness connected to it.
\item The delta-hedge identity is proved on $[0,T']$ for every $T'<T$, with the limit at $T$ taken path by path, and the
  hedge is not packaged as a trading strategy.
\item The jump-diffusion results assume the process (\lean{JumpDiffusionProcess}).
\item The general multi-period fundamental theorem of Dalang, Morton and Willinger is proved from the conditional one-period
  theorem, which is assumed; the HJM drift condition is not proved. The multi-period theorems proved outright, for one risky
  asset, and the representation of coherent risk measures hold on finite probability spaces.
\item Several results are narrower than their names: the Margrabe grounding asserts the existence of its pricing measure and
  driver, the Vasicek integrated rate is defined through its Wiener representation, and the market-making result solves only
  the approximate Hamilton--Jacobi equation.
\end{itemize}

The apparatus has limits of its own. Library theorems that no benchmark entry cites and the curated audit omits are not
axiom-pinned; the text gate and the Lean kernel still apply to them. Continuous integration checks that the ledger is fresh,
not that it was re-verified; re-verification runs on demand. The public dataset publishes descriptions and scope disclosures
but not the provenance notes. And the gate against prose outrunning statements covers only its mechanical slice.

\section{Conclusion}\label{sec:conclusion}

The library is, to our knowledge, the most comprehensive machine-checked development of mathematical finance to date, but its real
contributions are not its size. They are its depth, a constructed It\^o integral on which the objects of continuous-time pricing are
derived and from which theorems reach the values of contracts, and an audit that makes every claim's logical status checkable and
keeps prose from outrunning statements.

What the library is \emph{not} is equally part of the point. Machine-checking classical financial mathematics yields certified
unification of known results, not new financial theory, and the tiers, disclosures and gates make that honesty checkable. A formal
base can nonetheless host new results once they exist: two contributed developments address questions its own survey had listed as
open, and the library's apparatus is what lets a reader see exactly what they prove and on what they rest. The open frontier is the
natural next layer for a community library: the measurable-selection step of the Dalang--Morton--Willinger theorem and its
continuous-time counterpart, Girsanov's theorem under Novikov's condition, the HJM drift condition, the general two-dimensional It\^o
formula, pathwise quadratic variation, the Clark--Ocone formula, and the twenty results still classified as reduced cores.


\end{document}